\documentstyle[12pt,aps,epsf,amssymb]{revtex}
\begin{document}
\newcommand{\ini}{\begin{equation}}
\newcommand{\fin}{\end{equation}}
\newcommand{\inir}{\begin{eqnarray}}
\newcommand{\finr}{\end{eqnarray}}
\newcommand{\inif}{\begin{figure}}
\newcommand{\finf}{\end{figure}}
\newcommand{\bc}{\begin{center}}
\newcommand{\ec}{\end{center}}
\def\ol{\overline}
\def\pa{\partial}
\def\ra{\rightarrow}
\def\ts{{\times}}
\def\df{\dotfill}
\def\bs{\backslash}

$~$

\hfill DSF-T-99/36

\vspace{1 cm}

\centerline{\LARGE{Neutrino masses and mixings in $SO(10)$}}

\vspace{2 cm}

\centerline{\large{M. Abud$^*$, F. Buccella$^{*\circ}$,
D. Falcone$^{\circ}$, G. Ricciardi$^{\circ}$ and F.
Tramontano$^{*}$}}

\vspace{1 cm}

\centerline{$^*$Dipartimento di Scienze Fisiche, Universit\`{a} di Napoli,}
\centerline{Mostra d'Oltremare, Pad. 19, I-80125, Napoli, Italy,}
\centerline{and INFN, Sezione di Napoli, Italy;}
\centerline{$^{\circ}$Istituto di Fisica Teorica, Universit\`{a} di Napoli, Italy}

\vspace{2 cm}

\begin{abstract}

\noindent
Assuming a Zee-like matrix for the right-handed neutrino Majorana
masses in the see-saw mechanism, one gets maximal mixing for
vacuum solar oscillations, a very small value for $U_{e3}$ and an
approximate degeneracy for the two lower neutrino masses. The
scale of right-handed neutrino Majorana masses is in good
agreement with the value expected in a $SO(10)$ model with
Pati-Salam $SU(4)\ts SU(2)\ts SU(2)$ intermediate symmetry.

\end{abstract}

\newpage

\noindent
The evidence for neutrino oscillation, a phenomenon discussed
many years ago \cite{sol}, in solar \cite{exps} and atmospheric \cite{expa}
neutrinos, with square mass differences smaller than $1(eV)^2$,
has stimulated the search for theoretical framework \cite{thfr}. Unified
theories with $SO(10)$ as a gauge group \cite{so10} are a natural
choice, since there one expects neutrinos to be much lighter than
the other fermions as a consequence of the see-saw mechanism
\cite{ss}.

$SO(10)$ is a suitable framework to discuss fermion masses,
because all the left(right)-handed fermions of each family belong
to a single irreducible representation, namely the spinorial $\bf
{16~(\overline {16}})$. However, the spectrum and the mixing (the
$CKM$ matrix) of the fermions do not show a clear $SO(10)$
pattern. In fact, by assuming that the electroweak Higgs belongs
to a single real {\bf 10} representation, one gets the same
spectrum for the charge $\frac {2}{3}$ and $-\frac {1}{3}$ quarks
(and at the $SO(10)$ unification scale also for the Dirac
neutrino and charged lepton masses) and a trivial $CKM$ matrix.
This last prediction can be thought as a zero order approximation
in an expansion in $\lambda =\sin(\theta_c)$, but in order to
account for the large difference between $m_t$ and $m_b$, at
least one needs to assume that the two directions of the {\bf 10}
with vanishing colour and electric charge  have VEV's not
belonging to the same real {\bf 10} representations. In absence
of components of the electroweak VEV along representations higher
than {\bf 10} one keeps the equalities at the $SO(10)$ scale of
the quark with charge $-\frac {1}{3}$ and of the charged lepton
mass matrix, as well for the quark with charge $\frac {2}{3}$ and
the Dirac neutrino mass matrices. In that case one would have the
same mixings for the quark and the Dirac lepton mass matrix and,
especially for the two heaviest families, we can neglect the
mixing of Dirac leptons ($V_{cb}
\simeq 0.04$).

The effective Majorana mass matrix for the
light  left-handed neutrinos
is given by:

\ini
M_{\nu}^{(L)} = - M_{D}^{T}M_{\nu}^{(R)-1}M_{D}.
\fin
$M_{\nu}^{(R)}$ has to be symmetric and if we assume,
 for
simplicity, CP symmetry, it is a real matrix

 \inir
M_{\nu}^{(R)} = \left( \begin{array}{ccc} M_1 & \rho & \nu \\
\rho & M_2 & \mu \\ \nu & \mu & M_3,
\end{array} \right).
\finr

The Dirac mass matrix $M_D$ is assumed real and diagonal,
according to our approximation of neglecting the mixings:

\inir
M_D = \left( \begin{array}{ccc} M_{\nu_e}^{D} & 0 & 0 \\ 0 &
M_{\nu_\mu}^{D} & 0 \\ 0 & 0 & M_{\nu_\tau}^{D}
\end{array} \right).
\finr
From the previous equations it is easy to get

\inir
M_{\nu}^{(L)} = -\frac {1}{D} \left( \begin{array} {ccc}
\left( M_{2}M_{3}-{\mu}^2\right) \left(M_{\nu_e}^{D}\right)^2 &
\left(\mu\nu-M_{3}\rho\right)M_{\nu_e}^{D}M_{\nu_\mu}^{D} &
\left(\rho\mu-M_{2}\nu\right)M_{\nu_e}^{D}M_{\nu_\tau}^{D} \\
\left(\mu\nu-M_{3}\rho\right)M_{\nu_e}^{D}M_{\nu_\mu}^{D} &
\left(M_{1}M_{3}-{\nu}^2\right)\left(M_{\nu_\mu}^{D}\right)^2 &
\left(\rho\nu-M_{1}\mu\right)M_{\nu_\mu}^{D}M_{\nu_\tau}^{D} \\
\left(\rho\mu-M_{2}\nu\right)M_{\nu_e}^{D}M_{\nu_\tau}^{D} &
\left(\rho\nu-M_{1}\mu\right)M_{\nu_\mu}^{D}M_{\nu_\tau}^{D} &
\left(M_{1}M_
{2}-{\rho}^2\right)\left(M_{\nu_\tau}^{D}\right)^2
\end{array} \right)
\finr
where

\ini
D \equiv {\rm det} M_{\nu}^{(R)} =
 M_1M_2M_3 + 2\rho\nu\mu - M_1\mu^2 - M_2\nu^2 - M_3\rho^2.
\fin
Let us make the approximation of neglecting $M_{\nu_e}^D$ with
respect to the Dirac masses of the other neutrinos, which may be
reasonable if the electroweak doublet has components only along
{\bf 10} representations, which would imply

\ini
M_D = \frac {m_\tau}{m_b} M_{q = \frac {2}{3}},
\fin
but may hold more generally, since all the charged fermions of
the first family have masses smaller than those of the other two
families. In that limit the matrix elements of the first row and
of the first column of eq.(4) vanish and the remaining $2{\times}2$
matrix has the property of having a vanishing eigenvalue if one
takes only the contributions proportional to $\nu^2$, $\nu \rho$
and $\rho^2$. Since one of the solutions for the neutrino
spectrum is the hierarchical one with the highest mass $\simeq
\frac{1}{16}~eV$, we conclude that it is an intriguing possibility to have
only non-diagonal matrix elements in $M_{\nu}^{(R)}$. A matrix
with this property has been proposed many years ago by Zee
\cite{zee} for left-handed neutrinos and its phenomenological
consequences have been discussed by Frampton and Glashow
\cite{fg} in the case of light Majorana neutrinos. We
also assume
 a non-vanishing  $\mu$
in order to have $D \ne 0$. We define
\ini
A_e = \mu M_{\nu_e}^D = M_{0}^2 \sin \beta
\fin
\ini
A_{\mu}=\nu M_{\nu_{\mu}}^D = M_{0}^2 \cos \beta \cos
\alpha
\fin
\ini
A_{\tau}=\rho M_{\nu_{\tau}}^D = - M_{0}^2 \cos \beta \sin
\alpha
\fin
and get

$$
M_{\nu}^{(L)} = \frac{1}{D}\left( \begin{array}{ccc} A_{e}^2 &
-A_{e}A_{\mu} & -A_{e}A_{\tau} \\ -A_{e}A_{\mu} & A_{\mu}^2 & -A_{\mu}A_{\tau} \\
-A_{e}A_{\tau} & -A_{\mu}A_{\tau} & A_{\tau}^2
\end{array}\right)
$$
\ini
= \frac{M_{0}^4}{D} \left(
  \begin{array}{ccc}
    \sin^2 \beta & - \sin \beta \cos \beta \cos \alpha & \sin \beta \cos \beta \sin \alpha \\
    - \sin \beta \cos \beta \cos \alpha & \cos^2 \beta \cos^2 \alpha &
    \cos^2 \beta \cos \alpha \sin \alpha \\
    \sin \beta \cos \beta \sin \alpha & \cos^2 \beta \cos \alpha \sin \alpha &
    \cos^2 \beta \sin^2 \alpha
  \end{array} \right),
\fin
whose eigenvalues obey the secular equation

\ini
m_{i}^{3} -
m_{i}^{2}\frac{A_{e}^{2}+A_{\mu}^{2}+A_{\tau}^{2}}{D}+
\frac{4A_{e}^{2}A_{\mu}^{2}A_{\tau}^{2}}{D^3}=0\;\; \;\;\;\;\;
(i=1,2,3).
\fin
The absence of the linear term in eq.(11) implies for the three
solutions the relation

\ini
m_1 m_2 + m_1 m_3 + m_2 m_3=0.
\fin
If $m_3$ is larger than the other two solutions, one has that
$m_1$ and $m_2$ are almost opposite, more precisely
\ini
m_1+m_2=-\frac{m_1 m_2}{m_3}.
\fin
The ratios of the solutions of eq.(11) depend on the quantity

\ini
k^2=\frac{4A_{e}^{2}A_{\mu}^{2}A_{\tau}^{2}}{\left(A_{e}^{2}+
A_{\mu}^{2}+A_{\tau}^{2} \right)^3}=
\frac{1}{4} \sin^2 2\alpha \sin^2 2\beta \cos^2 \beta
\fin
which should be small in order to have one solution larger than the other
two. In this case one has approximately, to lowest order in $k$
\ini
m_3 \simeq
\frac{A_{e}^{2}+A_{\mu}^{2}+A_{\tau}^{2}}{D}=\frac{M_0^4}{D}
\fin
\ini
|m_1| \simeq |m_2| \simeq
\frac{2 A_e A_{\mu} A_{\tau}}{D \sqrt{A_{e}^{2}+A_{\mu}^{2}+A_{\tau}^{2}}}=
k \frac{M_0^4}{D}
\fin
and
\ini
m_1+m_2 \simeq \frac{4A_{e}^{2}A_{\mu}^{2}A_{\tau}^{2}} {D
\left(A_{e}^{2}+A_{\mu}^{2}+A_{\tau}^{2} \right)^2}=k^2 \frac{M_0^4}{D}
\fin
which imply
\ini
m_{1,2} \simeq k \frac{M_0^4}{D} \left(\mp1 + \frac{k}{2}
\right)
\fin
and
\ini
m_2^2 - m_1^2 = 2 k^3 \frac{M_0^8}{D^2}.
\fin
Under these assumptions,
$m_3^2$ and $m_2^2 - m_1^2$ have to be identified with
the values of $\Delta
m^2_{atm}$ and $\Delta m^2_{sun}$,
the quantities that are relevant for the oscillations
of the atmospheric and solar neutrinos. The parameter $k$
is so recovered to be
\ini
k^3 = \frac{1}{2} \frac{\Delta m^2_{sun}}{\Delta m^2_{atm}}.
\fin
The absence of the distortion
(which is predicted by the MSW solution)
of the spectrum, according to
 the new data of SuperKamiokande with a lower
threshold (5 MeV) for the electron detection, has brought Bilenky,
Giunti and Grimus \cite{bg2} to consider the vacuum solution
as the most probable with
the following parameters

\ini
\sin^2 2\theta_{atm}=1,~~\Delta m^2_{atm}=3.5\ts10^{-3}\, {\rm eV}.
\fin
\ini
\sin^2 2\theta_{sun}=0.8,~~\Delta m^2_{sun}=4.3\ts10^{-10}\, {\rm eV}.
\fin
From eq.(20),(21) and (22) we get
\ini
k=4\ts10^{-3}.
\fin
In order to have a small value for $k$, at least one of the $A$'s defined
in eqs.(7),(8) and (9) should be smaller than the others. The most
natural choice is $A_e$, since it is proportional to the Dirac
mass of $\nu_e$, that corresponds to a small value
for the $\beta$ angle. In this case $k \simeq
 \beta \sin 2 \alpha $, and
we are naturally brought to a small value for $U_{e3}$ and
to a large value
for the angle intervening in the solar neutrino oscillations.

In fact, the exact eigenvalues of eq.(11) are

\ini
m_3 = \frac{M^4_0}{D} \left( \frac{1}{3} + \frac{2}{3}\cos \psi
\right)
\fin
\ini
m_{2,1} = \frac{M^4_0}{D} \left[ \frac{1}{3} (1 -
\cos \psi) {\pm} \frac{\sqrt{3}}{3} \sin \psi \right]
\fin
with $\psi$ given by
\ini
\cos 3\psi = 1-\frac{27}{2} k^2.
\fin
The corresponding eigenvectors, in terms of $ \lambda_i = m_i
\frac{D}{M^4_0}$ are given by

\ini
a_i \, \, \left(
  \begin{array}{c}
    2 \lambda_i(\lambda_i-\cos^2 \beta) \\
    \sin 2 \beta \cos\alpha (-\lambda_i + 2\cos^2 \beta \sin^2 \alpha) \\
    \sin 2 \beta  \sin\alpha (\lambda_i - 2\cos^2 \beta \cos^2 \alpha) \\
  \end{array} \right)
\fin
where the $a_i$ are the appropriate normalization factors, to
have unit norm vectors. Developing in the small parameter
$\beta$, to  lowest order, we get:

\ini
|U_{e3}| \simeq \beta \cos 2\alpha \simeq k \cot 2\alpha
\fin
\ini
\sin^2 2\theta_{sun} =
\frac{4U^2_{e1}U^2_{e2}}{(|U_{e1}|^2 +
|U_{e2}|^2)^2} \simeq \left( 1- \frac{k^2}{4} \right) + O(k^3).
\label{sin}
\fin
For small $k$, the mixing angle for atmospheric neutrinos may be
identified with $\alpha$ and the high value for $\sin^2
2\theta_{atm}$ implies a very small value for $|U_{e3}|$, well
below the limit $|U_{e3}|^2 < 0.05$ \cite{bg1}, from the CHOOZ
experiment \cite{ch}. Eq. (\ref{sin}) implies that
 $\sin^2 2\theta_{sun}$ is practically equal to the
maximal value.

For $\alpha=\pi / 4$ one has, exactly to all orders,
 $U_{e3}=0$ and maximal mixing for
atmospheric neutrinos. In that case the mixing matrix, in the
limit $k=0$, is  the one with bimaximal mixing proposed in
ref.\cite{bim,bim2}

\inir
\left( \begin{array}{ccc} \frac{1}{\sqrt{2}} & \frac{1}{\sqrt{2}}
& 0 \\ -\frac{1}{2} & \frac{1}{2} & \frac{1}{\sqrt{2}} \\
-\frac{1}{2} & \frac{1}{2} & -\frac{1}{\sqrt{2}}
\end{array} \right).
\finr
A scenario similar to the one proposed here can be found in
ref.\cite{js} with an appropriate choice of the Dirac neutrino
mass and an antidiagonal form for the Majorana mass matrix for
right-handed neutrinos (in our notation with only $\nu$ and $M_2$
different from zero). Also in that paper the ratio of the scales
for solar and atmospheric neutrino oscillations are given in
terms of an expansion parameter in such a way that, when it
vanishes gives also rise to the bimaximal mixing (30). Also there
the two lower mass neutrino eigenstates are almost degenerate.
The resulting matrices for the two cases are different as well
the secular equation for the mass eigenstates, which brings to a
value for the mass of the almost degenerate lighter neutrinos
$\sqrt[3]{2}$ larger than the value given in ref.\cite{js}, a
difference which unfortunately seems very far from experimental
detection. Another difference concerns the matrix element
$M_{ee}$, which appears in the double $\beta$-decay, which is
predicted to vanish in \cite{js}, but as we shall see later, the
non-vanishing value predicted here is completely negligible for
the vacuum solution.

It is the right moment to remind that we have neglected the
rotation between the charged lepton and the Dirac neutrino mass
matrices. This approximation is reasonable if we assume that
electroweak Higgs has components only along {\bf 10}
representations of $SO(10)$ since in that case this solution is
 given by the CKM matrix, which has small non-diagonal
matrix elements. We are aware that the inequality $ m_\mu \, m_b
> m_{\tau} m_s $ requires some component of the electroweak Higgs
along some higher representation \cite{hrr}.

In $SO(10)$ the scale for the right-handed neutrino masses is
related to the scale of $B-L$ symmetry breaking \cite{bl} and
therefore it is interesting to see which values of $\mu$, $\nu$,
$\rho$ are needed to get the values of $\Delta
m^2_{atm}$ and $\Delta m^2_{sun}$. To the
purpose of getting the right order of magnitude, one can assume
eq.(6) and $\alpha =\pi/4$, which implies the value of $\mu$
\ini
\mu=\left( \frac{m_{\tau}}{m_b}\right)^2
\frac{m_c m_t}{m_3} \simeq 7.5 \ts 10^{11}~GeV.
\fin
We have  taken $m_3 = \sqrt{\Delta m^2_{atm}} = 6\ts 10^{-2} $~eV
and, as  reference values, the masses at the scale $M_Z$:
 $
m_u= 2.3 \ts 10^{-3}$~GeV, $m_c= 0.67 $~GeV, $m_b= 3.0 $~GeV,
$m_t= 181 $~GeV and $m_{\tau}= 1.75$~GeV \cite{fk}. Anyway, since
our quantities depend only on the ratios of the quark masses, a
strong dependence on the scale is not expected.

We also have

\ini
\nu=\mu \frac {m_u}{m_c} \frac{A_{\mu}}{A_e}=4.6 \ts 10^{11}
{\rm GeV},
\fin

\ini
\rho = \nu \frac{m_c}{m_t} = 1.7 \ts 10^{9} {\rm GeV}.
\fin

The rather moderate values for the matrix elements of
$M_{\nu}^{(R)}$ are a consequence of the approximate degeneracy
of the two lower neutrino mass eigenstate, which makes
\ini
m_{1,2} \simeq \sqrt{\frac{\Delta m_{sun}^2}{2}}
 \left( \frac{2 \Delta
m_{atm}^2}{ \Delta m_{sun}^2} \right)^{1/6} \gg \sqrt{\Delta
m_{sun}^2}
\fin
while with the hierarchical relationship $m_2 > m_1$ one should
have
\ini
m_1 \ll m_2 \simeq \sqrt{\Delta m_{sun}^2},
\fin
and the opposite inequality with respect to eq.(34). In fact from
the eqs.(1),(2),(5) and (6), and $m_3
\simeq \sqrt{\Delta m_{atm}^2}$, one would get
\ini
D = \frac{R^3 m_{u}^2 m_{c}^2 m_{t}^2}{m_{1} m_{2} m_{3}}
\simeq 2\ts 10^{35} \frac{\Delta m_{sun}^2}{m_1 m_2}~GeV^3
\fin
($R=(m_{\tau}/m_b)^2$) and the inequality (34) brings to a lower
value for D. Also the approximate degeneracy between $m_2$ and
$m_1$ allows to get  a not too broad
 neutrino mass spectrum.

For the first diagonal matrix elements of the Majorana mass one
finds as order of magnitude
\ini
M_{ee}=\frac {k m_u^2 \mu}{2 \rho \nu}\sim 1.2 \ts 10^{-6}eV;
\fin
much smaller than the present experimental limit
($0.2~eV$)\cite{beta}.

It is worth to recall that a value $\simeq 2.8\ts10^{11}~GeV$ has
been found \cite{buc} for the scale of spontaneous breaking of
$B-L$ in the $SO(10)$ model with $SU(4)\ts SU(2)\ts SU(2)$
intermediate symmetry \cite{ps}, which is broken by a VEV of the
{\bf 126} representation, endowed with the right quantum numbers
to give Majorana masses to the right-handed neutrino. Above that
scale, $SU(4)_{PS}$ for quarks and leptons implies that their
mass ratios do not change. The narrow range for the evolution of
$m_b/m_{\tau}$ is not enough to get $m_b=m_{\tau}$ at that scale \cite{acpr}
as required by the hypothesis that the electroweak Higgs VEV's
are only along {\bf 10} representations, but at least a smaller
contribution from other representation is needed with respect to
the case where the range of RGE for the mass ratio extends to the
unification scale, at which the lepto-quarks responsible for
proton decay take their mass.

As a conclusion we think that the choice of the Zee matrix
\cite{zee} for the right-handed neutrino Majorana masses seems
very appealing for the vacuum solution of the solar neutrino
problem and well consistent with the scale found in the $SO(10)$
model with Pati-Salam intermediate symmetry.

$~$

\noindent
One of us (F. B.) gratefully acknowledges stimulating discussions
with Profs. Z. Berezhiani and A. Masiero.


\begin{thebibliography}{100}

\bibitem{sol} B. Pontecorvo, JETP {\bf 53}, 1717 (1967);
V. Gribov and B. Pontecorvo, Phys. Lett. B {\bf 28}, 493 (1969).

\bibitem{exps} B.Pontecorvo, Report PD-205 National Research
Council of Canada,Division of Atomic Energy, Chalk River (1946).
R. Davis Jr $et~al.$, Phys. Rev. Lett. {\bf 20}, 1205 (1968);
Homestake Coll.: R. Davis Jr $et~al.$ Prog. Part. Nucl. Phys.
{\bf 32},13 (1994);
Gallex Coll.: P. Anselmann $et~al.$ Phys. Lett. B {\bf 342}, 440 (1995);
W. Hampel $et~al.$, Phys. Lett. B {\bf 447}, 127 (1999);
Sage Coll.: J. N. Abdurashitov $et~al.$ Phys. Lett. B {\bf 328}, 234 (1994);
Kamiokande Coll.: Y. Fukuda $et~al.$, Phys. Rev. Lett. {\bf 77}, 1683 (1996);
SuperKamiokande Coll.: Y. Fukuda $et~al.$ Phys. Rev. Lett.
{\bf 81}, 1158 (1998).

\bibitem{expa} Kamiokande Coll.: K.S.Hirata $et~al.$, Phys. Lett. B
{\bf 205}, 416 (1998); B {\bf 280}, 145 (1998);
M.B.Cooper $et~al.$, Phys. Rev. Lett. {\bf 66}, 2561 (1991);
R. Becker-Szendy $et~al.$, Phys. Rev. D {\bf 46}, 3720 (1992);
SuperKamiokande Coll.: Y. Fukuda $et~al.$ Phys. Rev. Lett.
{\bf 81}, 397 (1998);
Macro Coll.: M. Ambrosio $et~al$, Phys. Lett. B {\bf 343}, 451 (1998),
see also $Neutrino~98$, Nucl. Phys. B (Proc. Suppl.) {\bf 77} (1999).

\bibitem{thfr} J.K. Elwood, N. Irges and P. Ramond, Phys. Rev. Lett. {\bf 81},
5064 (1998);
K. Oda, E. Takasugi, M. Tanaka and M. Yoshimura, Phys. Rev. D {\bf 59}, 055001
(1999);
K. Hagiwara and N. Okamura, Nucl. Phys. B {\bf 548}, 60 (1999);
G. Altarelli and F. Feruglio, Phys. Lett. B {\bf 451}, 388 (1999);
Z. Berezhiani and A. Rossi, JHEP {\bf 03}, 002 (1999);
C.H. Albright, K.S. Babu and S.M. Barr, Phys. Rev. Lett. {\bf 81}, 1167 (1998);
J. Ellis, G.K. Leontaris, S. Lola and D.V. Nanopoulos, Eur. Phys. J. C {\bf 9},
389 (1999);
M. Fukugita, M. Tanimoto and T. Yanagida, Phys. Rev. D {\bf 59}, 113016 (1999);
H. Fritzsch and Z. Xing, Phys. Lett. B {\bf 440}, 313 (1998);
K. Matsuda, T. Fukuyama and H. Nishiura, hep-ph/9906433;
B. Stech, Phys. Lett. B {\bf 465}, 219 (1999).

\bibitem{so10} H. Georgi, in $Particles~and~Fields$ (AIP, New York, 1975);
H. Fritzsch and P. Minkowski, Ann. Phys. {\bf 93}, 193 (1975).
                                     
\bibitem{ss} M. Gell-Mann, P. Ramond and R. Slansky, in $Supergravity$,
eds. P. van Nieuwenhuizen and D. Freedman (North Holland,
Amsterdam, 1979);
T. Yanagida, in
$Proceedings~of~the~Workshop~on~Unified~Theory~and~Baryon~Number$
$in~the~Universe$, eds. O. Sawada and A. Sugamoto (KEK, 1979).

\bibitem{zee} A. Zee, Phys. Lett. B {\bf 93}, 389 (1980).

\bibitem{fg} P. H. Frampton and S. L. Glashow, Phys. Lett. B {\bf 461},
95 (1999). 

\bibitem{bg2} S. M. Bilenky, C. Giunti,
 W. Grimus, Prog. Part. Nucl. Phys. {\bf 43}, 1 (1999).
  
\bibitem{bg1} S. M. Bilenky and C. Giunti, Phys. Lett. B {\bf 444}, 379 (1998);
S.T. Petcov, hep-ph/9907216, to appear in the Proc. of Int.
 Workshop on Weak Interactions on Neutrinos (WIN 99).

\bibitem{ch} M. Apollonio $et~al.$, CHOOZ Collaboration,
 Phys. Lett. B {\bf 420}, 397 (1998).

\bibitem{bim} F. Vissani, hep-ph/9708483;
A. T. Baltz, A. S. Goldhaber and M. Goldhaber, Phys. Rev. Lett.
{\bf 81}, 5730 (1998);
V. Barger, S. Pakvasa, T. J. Weiler and K. Whisnant, Phys. Lett.
B {\bf 437}, 107 (1998).

\bibitem{bim2} Y. Nomura and T. Yanagida, Phys. Rev. D {\bf 59}, 017303 (1999);
G. Altarelli and F. Feruglio, Phys. Lett. B {\bf 439}, 112 (1998);
H. Georgi and S.L. Glashow, hep-ph/9808293;
S.K. Kang and C.S. Kim, Phys. Rev. D {\bf 59}, 091302 (1999);
C.H. Albright and S.M. Barr, Phys. Lett. B {\bf 461}, 218 (1999).

\bibitem{js} M. Jezabek and Y. Sumino, Phys. Lett. B {\bf 440}, 327 (1998).

\bibitem{hrr} J. Harvey, D.B. Reiss and P. Ramond, Nucl. Phys. B {\bf 199}, 223
(1982).

\bibitem{bl} D. Chang, R.N. Mohapatra, J. Gipson, R.E. Marshak
and M.K. Parida, Phys. Rev. D {\bf 31}, 1718 (1985); G. M. Gipson
and R. E. Marshak, Phys. Rev. D {\bf 31}, 1705 (1985);
F. Buccella, L. Cocco, A. Sciarrino and T. Tuzi, Nucl. Phys. B
{\bf 274}, 559 (1986);
F. Acampora, G. Amelino-Camelia, F. Buccella, O. Pisanti, L. Rosa
and T. Tuzi, Nuovo Cimento A {\bf 108}, 375 (1995).

\bibitem{fk} H. Fusaoka and Y. Koide, Phys. Rev D {\bf 57}, 3986 (1998).

\bibitem{beta} L. Baudis $et~al.$, Phys. Rev. Lett. {\bf 83}, 41 (1999).

\bibitem{buc} N. G. Deshpande, E. Keith and B. P. Pal,
 Phys. Rev. D {\bf 46}, 2261 (1992);
O. Pisanti and L. Rosa, Prog. Part. Nucl. Phys. {\bf 40}, 81
(1998);
F. Buccella and O. Pisanti, hep-ph/9910447.

\bibitem{ps} J. C. Pati and A. Salam, Phys. Rev. D {\bf 10}, 275 (1974).

\bibitem{acpr} H. Arason, D.J. Castano, E.J. Piard and P. Ramond, Phys. Rev. D
{\bf 47}, 232 (1993).

\end{thebibliography}
\end{document}